\documentclass[pre,twocolumn,showpacs,preprintnumbers,amsmath,amssymb]{revtex4}
\usepackage{bm,graphicx}
\begin{document}

\preprint{To appear in Phys.\ Rev.\ E (BR)}

\title{Exact calculations of the paranematic interaction energy for
colloidal dispersions in the isotropic phase of a nematogenic material}

\author{J.-B. Fournier}
\email{jbf@turner.pct.espci.fr}
\affiliation{Laboratoire de Physico-Chimie Th\'eorique,
E.\,S.\,P.\,C.\,I., 10 rue Vauquelin, F-75231 Paris cedex 05, France}
\author{P. Galatola}
\email{galatola@ccr.jussieu.fr}
\affiliation{LBHP, Universit\'e Paris 7---Denis Diderot, Case 7056, 2
place Jussieu, F-75251 Paris cedex 05, France}

\date{\today}

\begin{abstract}
In a recent paper [Phys.\ Rev.\ E \textbf{61}, 2831 (2000)], Bor{\v
s}tnik, Stark and {\v Z}umer have studied the stability of a colloidal
dispersion of micron-sized spherical particles in the isotropic phase of
a nematogenic material. Close to the nematic transition, the attraction
due to a surface-induced paranematic order can yield flocculation.
Their calculation of the nematic-mediated interaction was based on an
ansatz for the order-parameter profile. We compare it with an exact
numerical calculation, showing that their results are qualitatively
correct.  Besides, we point out that in the considered regime,
the exact interaction is extremely well approximated by a
simple analytical formula which is asymptotically exact.
\end{abstract}


\maketitle


In recent years, a large interest has been devoted to understanding the
interactions and phase behavior of colloidal particles dispersed in a
nematic phase~\cite{poulin_science97,poulin_prl97,stark99,%
yamamoto01,andrienko01} or in the isotropic phase of a nematogenic
compound~\cite{galatola99,borstnik99,borstnik00,galatola01}. In the
nematic phase, colloids experience a specific elastic interaction
because they induce competing distortions of the nematic director field.
New physics arises due to the long-range character of this interaction
and the induction of topological defects~\cite{poulin_science97}. 

In the isotropic phase, the surface of colloidal particles can induce a
local paranematic order~\cite{sheng76,miyano79}, giving rise to a
short-range elastic interaction~\cite{galatola99,borstnik99}. Two
effects compete: an \textit{attraction} due to the favorable overlapping
of the paranematic halos and a \textit{repulsion} due to the distortion
of the director field. For small particles, of size comparable to the
nematic--isotropic coherence length~$\xi$, it has been predicted that
repulsion may dominate and stabilize the colloidal
dispersion~\cite{galatola99,galatola01}. (Note that latex particles as
small as $50$\,nm have been successfully dispersed in lyotropic
nematics~\cite{mondain-monval99}.) On the other hand, Bor{\v s}tnik,
Stark and {\v Z}umer have predicted that for micron-sized particles
attraction dominates~\cite{borstnik99}, which allows to trigger
flocculation close to the nematic transition~\cite{borstnik00}.

The results of Bor{\v s}tnik, Stark and {\v Z}umer~\cite{borstnik00} are
based on a composite ansatz for the nematic director field
$\mathbf{n}$ and for the scalar order-parameter~$Q$, within an
uniaxial hypothesis. It turns out that our exact
calculations~\cite{galatola01}, based on a multipolar expansion for the
full tensorial order-parameter $Q_{ij}$, rest on the same
theoretical model, and can be performed also for micron-sized particles.

\begin{figure}
\includegraphics[width=.8\columnwidth]{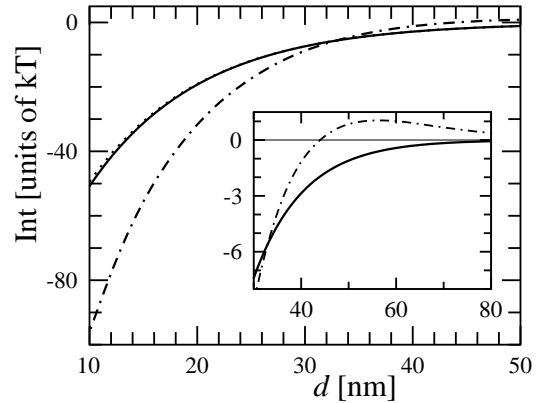}
\caption{\label{fig}
Paranematic interaction energy between two spherical particles of radius
$R=0.25\,\mu$m as a function of their distance to contact~$d$. The
parameters are $a=1.8\times10^5\,\mathrm{J\,m}^{-3}\,\mathrm{K}^{-1}$,
$T^*=313.5\, \mathrm{K}$, $\Delta T=1.3\,\mathrm{K}$,
$L_1=9\times10^{-12}\,\mathrm{J\,m}^{-1}$, $Q_s=0.3$,
$G_Q=G_n=10^{-3}\,\mathrm{J\,m}^{-2}$. The corresponding nematic
coherence length is $\xi=10.7$\,nm. The dashed-dotted line is extracted
from Fig.~7 of Ref.~\cite{borstnik99}. The full-line is the exact
result, numerically calculated according to Ref.~\cite{galatola01}.  The
dotted-line, practically coincident with the full-line, corresponds to
the asymptotic formula~(\ref{voila}).  } \end{figure}

In this Brief Report, we check the correctness of the paranematic
interaction energy used in Ref.~\cite{borstnik00}, by comparing it with
the exact one, numerically calculated according to the method of
Ref.~\cite{galatola01}. For the typical values considered in
Ref.~\cite{borstnik00}, we find that the exact interaction is attractive
instead of repulsive at distances $\agt5\xi$, and that it is about two
times weaker at distances of the order of~$\xi$ (Fig.~\ref{fig}).
However, the order of magnitude and the sign of the interaction is
correct in the regime where the paranematic attraction competes with the
electrostatic repulsion. In particular, at particles' separations of the
order of~$\xi$, the paranematic attraction remains much larger than the
van der Waals attraction: this implies that the conclusions of
Ref.~\cite{borstnik00} regarding critical flocculation phenomena remain
qualitatively correct. Finally, we show that in the \textit{whole regime
of interest}, the exact interaction calculated numerically is extremely
well approximated by a simple asymptotic formula [see Eq.~(\ref{voila})].

The Landau--de Gennes~\cite{deGennes_book} paranematic free-energy
density considered in Refs.~\cite{borstnik99,borstnik00} is
\begin{equation}
f=\frac{1}{2}a\,\Delta
T\,Q^2+\frac{3}{2}L_1\left(\bm{\nabla}Q\right)^2
+\frac{9}{2}L_1 Q^2 \left|\bm{\nabla}\mathbf{n}\right|^2
\end{equation}
for the bulk, and
\begin{equation}
\label{fstark}
f_s=G_Q\left(Q-Q_s\right)^2+3\,G_nQQ_s\sin^2\theta
\end{equation}
for the surface. Here $Q$ is the scalar order-parameter and $\mathbf{n}$
the nematic director. The parameter $Q_s$ is the order-parameter favored
by the particles' surface, and $\theta$ is the angle between the
direction of $\mathbf{n}$ at the surface and the normal to the
surface~$\bm{\nu}$, which is assumed to correspond to the easy-axis.
The coefficients $a$ and $L_1$ are material parameters, $\Delta
T=T-T^\star$ is the difference between the actual temperature~$T$ and
the limit of stability $T^\star$ of the isotropic phase, $G_Q$ and $G_n$
are introduced to describe the strength of the surface anchoring.

Incidentally, we note that the surface free-energy
density~(\ref{fstark}) is inconsistent within a Landau-de Gennes
framework, unless $G_Q=G_n$.  Indeed, at quadratic order, the most
general expansion in terms of the tensorial order parameter $Q_{ij}$ and
of the normal $\bm{\nu}$ to the surface can be written as
\begin{eqnarray}
f_s=&&g_1\,Q_{ij}\nu_i\nu_j+g_{21}\,Q_{ij}Q_{ij}
+g_{22}\,Q_{ij}Q_{jk}\nu_i\nu_k\nonumber\\*
&&+\,g_{23}\,Q_{ij}Q_{kl}\nu_i\nu_j\nu_k\nu_l,
\end{eqnarray}
where summation over repeated indices is implied.
For a uniaxial tensorial order-parameter
\begin{equation}
\label{uniax}
Q_{ij}=\frac{3}{2}Q\left(n_i n_j-\frac{1}{3}\delta_{ij}\right),
\end{equation}
this yields
\begin{eqnarray}
\label{fbon}
f_s=&&g_1 Q
+\frac{1}{2}\left[3\,g_{21}+2\left(g_{22}+g_{23}\right)\right]Q^2\nonumber\\*
&&-\,\frac{3}{4}\left[2\,g_1 Q+\left(g_{22}+4g_{23}\right)Q^2
\right]\sin^2\theta\nonumber\\*
&&+\frac{9}{4}g_{23}\,Q^2\sin^4\theta.
\end{eqnarray}
Matching Eqs.\ (\ref{fbon}) and~(\ref{fstark}) requires setting
$g_{23}=0$ and consequently $g_{22}=0$. Equation (\ref{fbon}) then
becomes
\begin{equation}
f_s=-\frac{{g_1}^2}{6\,g_{21}}
+\frac{3}{2}g_{21}\left[
\left(Q-Q_s\right)^2+3\,Q Q_s \sin^2\theta
\right],
\end{equation}
with $Q_s=-g_1/3g_{21}$. Therefore, Eq.~(\ref{fstark}) is compatible with
this expression only if $G_Q=G_n$.

With the required condition $G_Q=G_n\equiv G$, the free-energy
considered in Ref.~\cite{borstnik00} is identical to the one we used in
Ref.~\cite{galatola01}, with the correspondence~\footnote{Actually, in
our free-energy we keep the full biaxial order-parameter $Q_{ij}$
instead of assuming the uniaxial order~(\ref{uniax}).}: 
\begin{subequations}
\label{corresp}
\begin{eqnarray}
L^\dag&=&2 L_1,\quad
S^\dag=\frac{3}{2}Q,\quad
S_0^\dag=\frac{3}{2}Q_s,\\
a^\dag&=&\frac{2}{3}a\,\Delta T,\quad
W^\dag=\frac{4}{3}G,
\end{eqnarray}
\end{subequations}
where we have dagged the quantities appearing in Ref.~\cite{galatola01}.
With the above relationships, we have numerically recalculated the
exact interaction energy between two spherical particles of radius~$R$
as a function of their distance to contact~$d$---using the same
parameters as in Ref.~\cite{borstnik99}. The comparison with the results
of the ansatz of Bor{\v s}tnik, Stark and {\v Z}umer~\cite{borstnik99}
is shown in Fig.~\ref{fig}. We find a qualitative agreement, as
previously discussed. Note that the exact nematic director profile
displays a Saturn-ring defect~\cite{galatola01}, which is absent in the
ansatz of Ref.~\cite{borstnik99}. For the case of micron-sized particles
considered in Refs.~\cite{borstnik99,borstnik00}, this defect lies
however in a region where the paranematic is almost completely melted.

Finally, we have compared our numerical calculation with the asymptotic
formula obtained by us in Ref.~\cite{galatola01}.  With
the correspondence~(\ref{corresp}), the latter reads
\begin{equation}\label{voila}
F=-48\pi L_1\xi\left(\frac{Q_s}{A}\right)^2\frac{e^{\displaystyle-\bar
d}}{2\bar R+\bar d}\,,
\end{equation}
where $\bar d=d/\xi$, $\bar R=R/\xi$, and
$A$ is a constant given by
\begin{equation}
A=\frac{27\bar\ell}{\bar R^4}
+\frac{6+27\bar\ell}{\bar R^3}
+\frac{6+12\bar\ell}{\bar R^2}
+\frac{2+3\bar\ell}{\bar R},
\end{equation}
where $\bar\ell=L_1/G\xi$ is the reduced extrapolation length of the
anchoring. The nematic coherence length is
$\xi=\sqrt{3L_1/a\,\Delta T}$. As shown in Fig.~\ref{fig}, the agreement
between the numerical and the analytical calculations is excellent in
the range of separations relevant to the colloidal flocculation
discussed in Ref.~\cite{borstnik00}. Owing to its simplicity and
validity, formula~(\ref{voila}) thus offers a straightforward means to
systematically investigate the stability of such paranematic-wetted
colloids.



\begin{thebibliography}{13}
\expandafter\ifx\csname natexlab\endcsname\relax\def\natexlab#1{#1}\fi
\expandafter\ifx\csname bibnamefont\endcsname\relax
  \def\bibnamefont#1{#1}\fi
\expandafter\ifx\csname bibfnamefont\endcsname\relax
  \def\bibfnamefont#1{#1}\fi
\expandafter\ifx\csname citenamefont\endcsname\relax
  \def\citenamefont#1{#1}\fi
\expandafter\ifx\csname url\endcsname\relax
  \def\url#1{\texttt{#1}}\fi
\expandafter\ifx\csname urlprefix\endcsname\relax\def\urlprefix{URL }\fi
\providecommand{\bibinfo}[2]{#2}
\providecommand{\eprint}[2][]{\url{#2}}

\bibitem[{\citenamefont{Poulin et~al.}(1997{\natexlab{a}})\citenamefont{Poulin,
  Stark, Lubensky, and Weitz}}]{poulin_science97}
\bibinfo{author}{\bibfnamefont{P.}~\bibnamefont{Poulin}},
  \bibinfo{author}{\bibfnamefont{H.}~\bibnamefont{Stark}},
  \bibinfo{author}{\bibfnamefont{T.~C.} \bibnamefont{Lubensky}},
  \bibnamefont{and} \bibinfo{author}{\bibfnamefont{D.~A.} \bibnamefont{Weitz}},
  \bibinfo{journal}{Science} \textbf{\bibinfo{volume}{275}},
  \bibinfo{pages}{1770} (\bibinfo{year}{1997}{\natexlab{a}}).

\bibitem[{\citenamefont{Poulin et~al.}(1997{\natexlab{b}})\citenamefont{Poulin,
  Cabuil, and Weitz}}]{poulin_prl97}
\bibinfo{author}{\bibfnamefont{P.}~\bibnamefont{Poulin}},
  \bibinfo{author}{\bibfnamefont{V.}~\bibnamefont{Cabuil}}, \bibnamefont{and}
  \bibinfo{author}{\bibfnamefont{D.~A.} \bibnamefont{Weitz}},
  \bibinfo{journal}{Phys. Rev. Lett.} \textbf{\bibinfo{volume}{79}},
  \bibinfo{pages}{4862} (\bibinfo{year}{1997}{\natexlab{b}}).

\bibitem[{\citenamefont{Stark et~al.}(1999)\citenamefont{Stark, Stelzer, and
  Bernhard}}]{stark99}
\bibinfo{author}{\bibfnamefont{H.}~\bibnamefont{Stark}},
  \bibinfo{author}{\bibfnamefont{J.}~\bibnamefont{Stelzer}}, \bibnamefont{and}
  \bibinfo{author}{\bibfnamefont{R.}~\bibnamefont{Bernhard}},
  \bibinfo{journal}{Eur. Phys. J. B} \textbf{\bibinfo{volume}{10}},
  \bibinfo{pages}{515} (\bibinfo{year}{1999}).

\bibitem[{\citenamefont{Yamamoto}(2001)}]{yamamoto01}
\bibinfo{author}{\bibfnamefont{R.}~\bibnamefont{Yamamoto}},
  \bibinfo{journal}{Phys. Rev. Lett.} \textbf{\bibinfo{volume}{87}},
  \bibinfo{pages}{075502} (\bibinfo{year}{2001}).

\bibitem[{\citenamefont{Andrienko et~al.}(2001)\citenamefont{Andrienko,
  Germano, and Allen}}]{andrienko01}
\bibinfo{author}{\bibfnamefont{D.}~\bibnamefont{Andrienko}},
  \bibinfo{author}{\bibfnamefont{G.}~\bibnamefont{Germano}}, \bibnamefont{and}
  \bibinfo{author}{\bibfnamefont{M.~P.} \bibnamefont{Allen}},
  \bibinfo{journal}{Phys. Rev. E} \textbf{\bibinfo{volume}{63}},
  \bibinfo{pages}{041701} (\bibinfo{year}{2001}).

\bibitem[{\citenamefont{Galatola and Fournier}(1999)}]{galatola99}
\bibinfo{author}{\bibfnamefont{P.}~\bibnamefont{Galatola}} \bibnamefont{and}
  \bibinfo{author}{\bibfnamefont{J.-B.} \bibnamefont{Fournier}},
  \bibinfo{journal}{Mol. Crys. Liq. Cryst.} \textbf{\bibinfo{volume}{330}},
  \bibinfo{pages}{535} (\bibinfo{year}{1999}).

\bibitem[{\citenamefont{Bor{\v s}tnik et~al.}(1999)\citenamefont{Bor{\v s}tnik,
  Stark, and {\v Z}umer}}]{borstnik99}
\bibinfo{author}{\bibfnamefont{A.}~\bibnamefont{Bor{\v s}tnik}},
  \bibinfo{author}{\bibfnamefont{H.}~\bibnamefont{Stark}}, \bibnamefont{and}
  \bibinfo{author}{\bibfnamefont{S.}~\bibnamefont{{\v Z}umer}},
  \bibinfo{journal}{Phys. Rev. E} \textbf{\bibinfo{volume}{60}},
  \bibinfo{pages}{4210} (\bibinfo{year}{1999}).

\bibitem[{\citenamefont{Bor{\v s}tnik et~al.}(2000)\citenamefont{Bor{\v s}tnik,
  Stark, and {\v Z}umer}}]{borstnik00}
\bibinfo{author}{\bibfnamefont{A.}~\bibnamefont{Bor{\v s}tnik}},
  \bibinfo{author}{\bibfnamefont{H.}~\bibnamefont{Stark}}, \bibnamefont{and}
  \bibinfo{author}{\bibfnamefont{S.}~\bibnamefont{{\v Z}umer}},
  \bibinfo{journal}{Phys. Rev. E} \textbf{\bibinfo{volume}{61}},
  \bibinfo{pages}{2831} (\bibinfo{year}{2000}).

\bibitem[{\citenamefont{Galatola and Fournier}(2001)}]{galatola01}
\bibinfo{author}{\bibfnamefont{P.}~\bibnamefont{Galatola}} \bibnamefont{and}
  \bibinfo{author}{\bibfnamefont{J.-B.} \bibnamefont{Fournier}},
  \bibinfo{journal}{Phys. Rev. Lett.} \textbf{\bibinfo{volume}{86}},
  \bibinfo{pages}{3915} (\bibinfo{year}{2001}).

\bibitem[{\citenamefont{Sheng}(1976)}]{sheng76}
\bibinfo{author}{\bibfnamefont{P.}~\bibnamefont{Sheng}},
  \bibinfo{journal}{Phys. Rev. Lett.} \textbf{\bibinfo{volume}{37}},
  \bibinfo{pages}{1059} (\bibinfo{year}{1976}).

\bibitem[{\citenamefont{Miyano}(1979)}]{miyano79}
\bibinfo{author}{\bibfnamefont{K.}~\bibnamefont{Miyano}},
  \bibinfo{journal}{Phys. Rev. Lett.} \textbf{\bibinfo{volume}{43}},
  \bibinfo{pages}{51} (\bibinfo{year}{1979}).

\bibitem[{\citenamefont{Mondain-Monval
  et~al.}(1999)\citenamefont{Mondain-Monval, Dedieu, Gulik-Krzywicki, and
  Poulin}}]{mondain-monval99}
\bibinfo{author}{\bibfnamefont{O.}~\bibnamefont{Mondain-Monval}},
  \bibinfo{author}{\bibfnamefont{J.~C.} \bibnamefont{Dedieu}},
  \bibinfo{author}{\bibfnamefont{T.}~\bibnamefont{Gulik-Krzywicki}},
  \bibnamefont{and} \bibinfo{author}{\bibfnamefont{P.}~\bibnamefont{Poulin}},
  \bibinfo{journal}{Eur. Phys. J. B} \textbf{\bibinfo{volume}{12}},
  \bibinfo{pages}{167} (\bibinfo{year}{1999}).

\bibitem[{\citenamefont{de~Gennes and Prost}(1993)}]{deGennes_book}
\bibinfo{author}{\bibfnamefont{P.-G.} \bibnamefont{de~Gennes}}
  \bibnamefont{and} \bibinfo{author}{\bibfnamefont{J.}~\bibnamefont{Prost}},
  \emph{\bibinfo{title}{The Physics of Liquid Crystals}}
  (\bibinfo{publisher}{Clarendon}, \bibinfo{address}{Oxford},
  \bibinfo{year}{1993}).

\end{thebibliography}
\end{document}